\documentclass{./llncs}
\usepackage{makeidx}  
\usepackage[english]{babel}
\pdfoutput=1
\usepackage[utf8]{inputenc}

\usepackage{multicol}
\usepackage{boxedminipage}

\usepackage{graphicx}
\usepackage{fancybox}
\usepackage{pdfpages}

\usepackage{amsmath}
\usepackage{caption}
\usepackage{diagrams}
\usepackage{alltt}

\usepackage{boxedminipage}
\usepackage{multicol}
\usepackage[pdftex]{thumbpdf}
\usepackage[pdfusetitle]{hyperref}
\usepackage{cleveref}
\usepackage{tikz}

\usepackage{xcolor}
\usepackage{todonotes}
\usepackage{menukeys}

\presetkeys{todonotes}{color=green!20}{}

\newcommand{\red}[1]{\color{red}  #1}
\newcommand{\blue}[1]{\color{blue}  #1}
\newcommand{\selectP}{\uparrow} 

\begin{document}
%
%
\pagestyle{headings}  

%
%
%
\title{Mastering Heterogeneous Behavioural Models}

\titlerunning{running title}  
%
\author{J. Christian Attiogb{\'e} \\[0.7ex]  
}
\institute{LS2N  - UMR CNRS 6004 - University of Nantes\\
  \email{Christian.Attiogbe@univ-nantes.fr}\\ $~~$ \\
  \textit{A detailed version of a paper accepted in the MEDI'2017 conference}
}


\maketitle              

\begin{abstract}
Heterogeneity is one important feature of complex systems, leading to the complexity of their construction and analysis. Moving the heterogeneity at model level helps in mastering the difficulty of composing heterogeneous models which constitute a large system. 
We propose a method made of an algebra and structure morphisms to deal with the interaction of behavioural models, provided that they are compatible. 
We prove that heterogeneous models can interact in a safe way, and therefore complex heterogeneous systems can be built and analysed incrementally. The \textsf{Uppaal} tool is targeted for experimentations.
\end{abstract}

\section{Introduction}
\label{section:intro}
%

Mastering the composition of heterogeneous models contributes to settle the challenge of building and analyzing large systems. The subsequent benefits are the gain of flexibility and an easier evolution of systems construction.

Models are used at different abstraction levels and for different purposes. Data models (also qualified as static models)  capture  the structure of manipulated data; behavioural models (also called dynamic models) often based on transition systems or event systems help to predict and reason about the  behaviour of the software to be built.
Other models such as timed models, security models, functional models are required according to the needs.
A combination of these models is often necessary. In this article we focus on behavioural models often used to capture the evolution and the interaction between parts (called processes) of a more general heterogeneous system which can combine hardware and various software components.

An example of an heterogeneous system is an assembly of pieces of software and hardware communicating with a distributed architecture (using smart objects, sensors, actuators, mechanical parts driven by software). This kind of systems is spreading more and more as users adopt their provided services and practical facilities. 
However mastering their design,  proving their correctness and maintaining these systems are a challenge of first importance for the security of services and software, and especially for reducing time to market of smart objects.

The integration of various formalisms inside a main one is often used as a solution to master the difficulties of heterogeneous systems. This constitutes an approach with a strong coupling of the formalisms; we dot not follow this trends, since we consider that a weak coupling favours reuse and flexibility.

There are several works and proposals related to heterogeneous issues; they embrace different abstraction levels and adopt various policies. Interface Theories are intensively studied \cite{alfarohezinger2001interface}; they are based on reasoning on the contracts that link components and the information provided in their interfaces. 
SysML \cite{SysML-Friedenthal2015} adresses system engineering at modelling levels.
SystemC \cite{systemC2010} adopts a rather low abstraction level by composing software or hardware modules which are classes containing processes modelling functionalities. The core of SystemC  consists of an event-driven simulator working as a scheduler. 
The Ptolemy project \cite{GoderisBALG09,DBLP:journals/pieee/EkerJLLLLNSX03} proposes one of the most advanced framework, Ptolemy II \cite{ptolemy-TripakisSSL13} with which we share some concerns. But these are  general purpose and heavy weight approaches which, from our point of view, constraint a lot the used components; they build a kind of a scheduler of the whole composition of components. We rather target a specific framework with a very weak coupling of components that can be composed (plugged) or unplugged at any time.
For that, we consider the behaviours of the interacting components described with different formalisms, hence the heterogeneity. We especially focus on the communication channels that they use.
We address systems with evolving adhoc structures, for small aperture net of components. With our approach the designer, before going further on development or during system maintenance, can for instance check if plugging a new component leads to a correct interaction or not.  



This work is motivated by the necessity of light methods and tools to face the construction and the analysis of heterogeneous systems; these methods and tools should ensure flexibility, easy of use and evolution.
The difficulties of heterogeneity arise not only at language level (data, property or behavioural), but also at the semantic level. The latter being more challenging. 

We propose a method to make it easier the composition and hence the interaction between heterogeneous but compatible behavioural models. The idea is that one can easily compose models described with different formalisms but having the same compatibility domain, for instance their semantics are translated into transition systems. Currently we focus on such behavioural models. We implement a prototype tool, namely \textsf{aZiZa}, to support the experimentations with the method.

The article is organized as follows. 
Section \ref{section:materials} introduces the materials we have used.
Section \ref{section:interaction} is devoted to the proposed method, an algebra to structure the composition of models.
Section \ref{section:experimentation} deals with a complete experimentation supported by the developed tools.
Section  \ref{section:conclu} concludes the article.

\vspace{-0.3cm}
\section{Materials: Models Compatibility and Composition}
 \label{section:materials}

Two main features have to be distinguished in terms of modelling: description formalisms and semantic models.
First order logic, grammars,  automata are examples of description formalisms.
Decision tables, labelled transition systems, operational semantic rules, axiomatic systems are examples of semantic models.

Heterogeneity is mastered at both levels; but we focus on semantic models which we consider as  \textit{compatibility domains}. 
Several categories of compatibility domains can be considered, for instance
labelled transition systems,
event-based models,
predicate transformer \textit{à la Dijkstra}.

We consider mainly the semantic models used by the modelling formalisms and extend them to the mathematical foundations they use.
Automata and processes, as description formalisms, share the same semantic domain: Labelled Transition Systems (LTS).

\subsubsection*{Labelled Transition Systems}
Given a set of states $S$, a set of labels $L$, a labelled transition system is defined by the tuple  $\langle S, L, \stackrel{l}{\to} \rangle$ where $\stackrel{l}{\to}$ is a transition relation: $\stackrel{l}{\to}  : S \times L \rightarrow S$

\subsubsection*{Product of Transition Systems}
\label{subsection:freeProduct}
The parallel composition of processes, can be built by the \textit{free product} of the transition systems of the processes. Theoretically, their free product results in a  model where a global state\footnote{The term \textit{global state} refers to the state of the process composition} is made of a state of each of the processes, a global transition from a global state is made of the transitions of each of the processes from its state involved in the global state.  However, practically, the global transitions can be constrained to avoid inconsistency in the access or update of the state variables.

In general free product is appropriate to the composition of independent (asynchronous) processes which may share communication channels.
Besides, the \textit{synchronous product} is appropriate to highly synchronous processes.

In our work, the considered global systems are asynchronous and distributed; their components can be concurrent and dynamic and the communication channels can be synchronous or asynchronous.


\vspace{-0.2cm}
\subsubsection*{Compatible Models}
Two models $M_1$ and $M_2$ (or more) are said  \textit{compatible} or not, with regards to at least three compatibility levels:
syntactic compatibility,
semantic compatibility and
formal-reasoning compatibility.

Syntactic compatibility can be solved at description formalisms level with (syntactic) model transformations known as \textit{shallow embedding} \cite{GordonHOL}; 
semantic compatibility permits a \textit{deep embedding} \cite{GordonHOL} of one model into the other provided that there are a compatibility domain: when the semantics of one model can be expressed in terms of the semantics model underlining the other model;
the semantic compatibility is also used as the basis for formal reasoning.\\

From the formal point of view,  first order logics and higher order logics provide logical frameworks with powerful constructs to describe and reason on heterogeneous objects, whatever their nature.\\
Transition Systems  \cite{TS_Arnold93}, Mealy Machines \cite{RothKinney_Mealy_2004}, with their various extensions, are widely  used to handle complex dynamic systems and are at the heart of many analysis methods and verification tools. The underlying theories are well-studied and, in the current state of the art a lot of effective systems are \textit{compiled} into transition systems. 
Process Algebra (such as CCS \cite{Mil89}, CSP\cite{Roscoe98,DBLP:RoscoeD11}, LOTOS\cite{ISOlotos88}, $\pi$-calculus \cite{Milner92}) built on top of transition systems are recognized as powerful behavioural description models; they are also representative of many behavioural languages, hence their use as composition and interoperability basis.

\begin{definition}(Compatibility Domain)
A \textit{Compatibility Domain} is defined as a category of models characteristics  in such a way that, any two models considered within this domain, are comparable w.r.t. the considered characteristics. It is a model integration basis. 
\end{definition}

Examples of \textit{semantic compatibility domains} are: logics, labelled  transition systems, trace semantics, temporal logics, weakest preconditions, Kripke model, algebra. For instance first order logic enables reasoning on several formalisms which use logic for objects description; when object descriptions (from different formalisms) have been translated as predicates in first order logic, the resulting predicates can be combined by the operators of first order logic.
Examples of \textit{syntactic compatibility domains} are grammars, set-theory, flowcharts.

Compatibility domains are the basic frameworks to compose or to integrate models within a global system.

\begin{proposition} Within a compatibility domain, it is always possible to translate objects semantics (from one formalism and paradigm) into the domain, to compose or integrate them, to reason within the basis, and to possibly translate results in  target formalisms.
\end{proposition}

\medskip
The practical use in our work of compatibility domain is the definition of bridges between models and their semantics.
We need the definition of \textit{model bridges} between models in syntactical and semantic levels. 
Roughly, the principle of a bridge is to embed the model underlying a description into a given semantic model which will be used at model composition level.

\subsubsection{Semantic Models and Semantic Embedding}
A direct application of the notion of compatibility is the construction of semantic bridges between models or the semantic embedding of one model into another one.  
We choose the LTS as a reference behavioural model, because it is widely used and equipped with various tools. 

If two models $M_1$ and $M_2$ are in a compatible semantic domain (LTS in our case), it exists  structure morphisms $\zeta_1$ and $\zeta_2$ with related meaning matching  such that $\zeta_1(M_1) = LTS_1$ and $\zeta_2(M_2) = LTS_2$.

Accordingly, we are about to define some operators $\Phi$ which arguments are different but compatible behavioural models; these operators form an algebra that leads the interaction of behavioural models. The idea is that $\Phi(M_i, M_j)$ is semantically unfolded as $\phi(\zeta_i(M_j),\zeta_j(M_j))$ where $\phi$ is the domain-compatible equivalent of $\Phi$.


If we consider a behavioural model $M$ as a term of a given algebra ${\mathcal A}$, a sketch of the semantic embedding of models is as follows; we consider  ${\mathcal A_i}$ as the source algebra to describe various but compatible models, then it exists a compatible domain denoted here by $({\mathcal S}, L, \rightarrow)$ the LTS.

\begin{diagram}
{\red \Phi~level} ~~~~~~~ & {\mathcal A_1}   & ~ ~ . ~ ~ & {\mathcal A_i} & \cdots & {\mathcal A_j} & ~ ~ . ~ ~ & {\mathcal A_n}\\
 &  & ~ \rdTo(3,2)^{\zeta_1} & & \rdTo(1,2)^{\zeta_i}  \ldTo(1,2)^{\zeta_j} & & \ldTo(3,2)^{\zeta_n} \\
{\red \phi~level} ~~~~~~~ &  & ~ ~  ~ ~ & ~ ~ &   ({\mathcal S}, L, \rightarrow) & ~ ~ ~ ~ &  &  \\
\end{diagram}


It follows that when models are compatible, a bridge can be used to relate them via the semantic structure induced by the compatibility domain (for instance their LTS). Consequently, a multilevel bridge can be gradually built between compatibility domains to link two or more models.

\vspace{-0.3cm}
\section{Interaction of Hererogeneous  Models: an Algebra}
\label{section:interaction}
%

Interaction between behavioural models, whatever their description formalisms, is viewed as exchanges through common communicating channels. Typically the interaction is denoted by a flow of emission and reception statements.
Process algebra models, as a compatibility domain, capture very well these interactions, where the unit of specifications is a \textit{process} expressing an elementary sequential behaviour; more complex behaviours are expressed with the composition (sequential, parallel, etc)  of other processes, elementary or not. 

Handling the heterogeneity is as simpler as if the LTS is the known user manual of each component. On the one hand, we extract the LTS from  given components to compose them; on the other hand the LTS can be given by the component providers. Besides, an implementation can be built from a LTS used to tune a composition.

\begin{figure}[!ht]
\begin{center}
\includegraphics[width=0.5\linewidth]{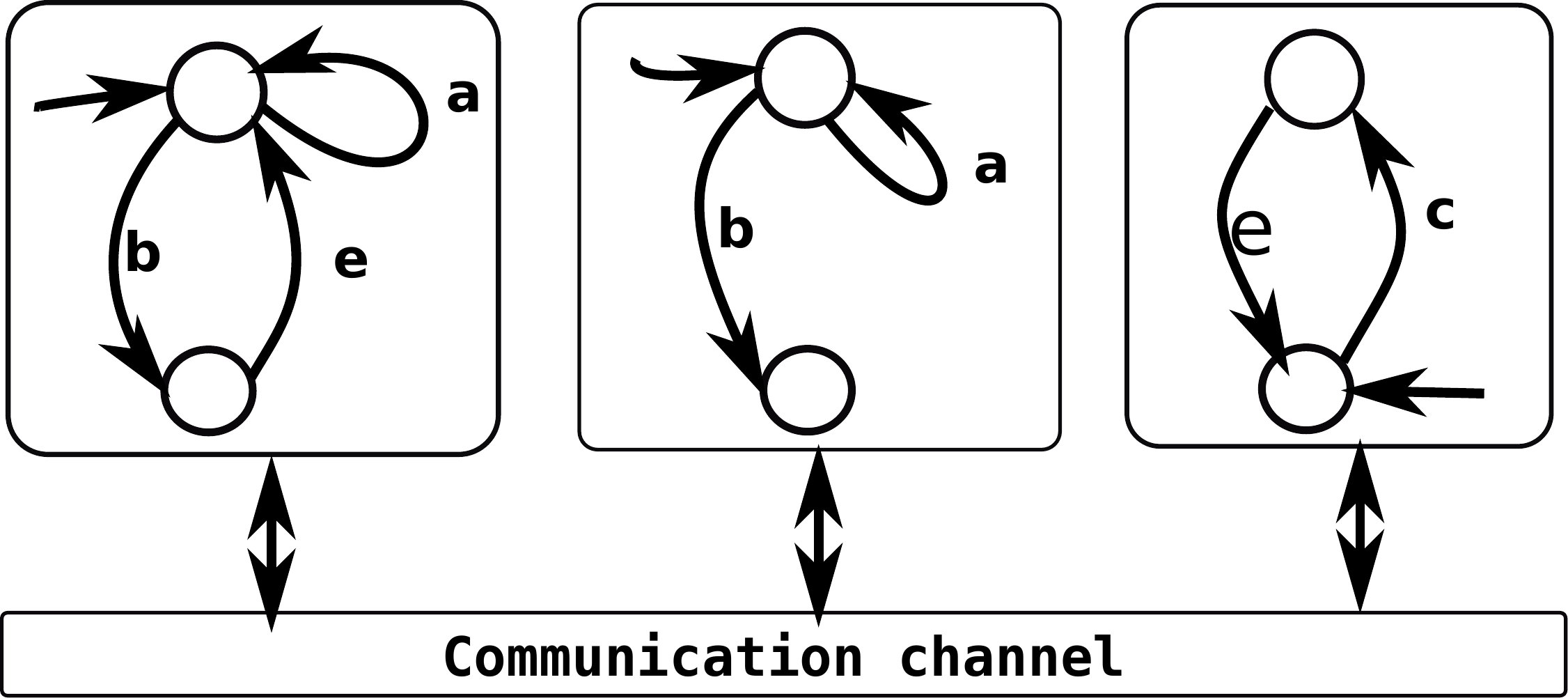}
 \caption{Interaction between the processes: model level}
\label{scheme3processesA}
\end{center}
\end{figure}

Therefore an abstraction of the communicating processes can be considered (see Fig.\ref{scheme3processesA}) where the heterogeneity of the source models is tackled via the abstract behaviours captured with transition systems. According to Fig.\ref{scheme3processesA}, if $a$, $b$ and $e$ are synchronous actions, then the interacting processes can for instance perform the trace:  \textsf{c.a*.b.e}. We only need the communication actions of each process, not the entire behaviour. However, we must make precise the semantics of this interaction, and generalize the communication hypothesis; typically the relationship between synchronisation actions, and communication channels.

\noindent We define a set of operators that impact the behaviour of process composition:
\begin{itemize}
\item a process composition can be restructured through the  renaming of channels;
\item process communications can be broken through the modification of channels; 
\item the structure of a complete net of processes can evolve through channel restructuring, etc
\end{itemize}

\vspace{-0.2cm}
\subsection{The Core Operators for Model Interaction}

We are about to elaborate an algebra  ${\mathcal A}$ to structure and analyse the composition of heterogeneous processes. The operators of the algebra are related to the two levels ($\Phi$ level and $\phi$ level) considered in Section~\ref{section:materials}, while the structure set of the algebra is the set of processes ${\mathcal P}$ to be composed. Our target is an algebra ${\mathcal A} = \langle {\mathcal P},  O_{\Phi}, O_{\phi} \rangle$; therefore we introduce these sets of operators.\\

In the following, a \textit{process} is denoted by the term:

\centerline{\textsf{Process procName [channel parameters](other parameters) \{body\}}}

where we consider its name, its channels and parameters, and a body. The body is an LTS which describes the behaviour of the process. Several named instances of a process can be defined using the process name as a type. 

A \textit{system} is made with the composition of at least two processes.

\vspace{-0.2cm}
\paragraph{\textbf{\textsf{compose}: abstract parallel composition of processes}} 
Let $P_1$ and $P_2$ be two processes which use a shared communication channel \textsf{nc}.
\begin{center}
\textsf{Process Proc[nc]() P1},\\
\textsf{Process Proc[nc]() P2}.
\end{center}

The expression $S = \textsf{compose}(P_1, P_2)$ is the system made of the parallel composition of the processes $P_1$ and $P_2$ which interact via their common \textsf{nc} channel. Note that a channel can be hidden in a process by the renaming of the channel. 
The arity of the \textsf{compose} operator is not a strong constraint; a set of $n$ processes can be composed either with the binary composition\\
\centerline{$\textsf{compose}(\dots(\textsf{compose}(\textsf{compose}(P_1, P_2), P_3),\cdots),P_n)$}
or directly with the list of processes as arguments:
$\textsf{compose}(P_1, P_2, \cdots, P_n)$.

The \textsf{compose} operator is an instance of the $\Phi$ operator. Typically, the embedding  functions $\zeta_i$ compute the transition systems from the processes used as arguments of \textsf{compose}; then $\phi$ is the synchronous product \cite{DBLP:RoscoeD11} of the resulting transition systems.

A component process of a system built by the \textsf{compose} operator may be \textbf{selected} with the projection operator denoted by $\selectP$. Consequently an operation $\alpha$ can be applied to a process inside a composition by selecting it as follows: $\alpha(\textsf{compose}(P_1, P_2, \cdots, P_n)\selectP P_3)$.

\vspace{-0.2cm}
\paragraph{\textbf{\textsf{rename}: renaming a channel in a process}}
The expression ($P$ \textsf{rename}~ $c$~ \textsf{as}~ $nc$) denotes a process $P$ where the channel $c$ is renamed as $nc$. 

Let $P_3$ be a process using $nc$ as a channel: \textsf{Process Proc[nc]() P3}. 
The expression $S = \textsf{compose}(P_1, P_2~\textsf{rename}~ nc~ \textsf{as} ~c, P_3)$ results in a system where only $P_1$ and $P_3$ interact through $nc$. The behaviour of $P_2$ does not impact the behaviour of $S$ since $P_2$ uses a local channel, thus the behaviour of $P_2$ is ignored in $S$.



\vspace{-0.2cm}
\paragraph{\textbf{\textsf{replace}: substitution of processes}}
Within a system, a given process is substituted by a given new one. The  \textsf{replace} operator needs three arguments: a system $S$, a process $oP$ already in $S$, a new process $nP$ not in $S$. The process $oP$ should share its channels with $S$. The process $nP$ should have the same shared channels (for the substitution) but it can have more channels.

The effect of the replacement is  based on the shared channels; the shared channel $oP$  is cut and replaced by the common channel in $nP$.  
The expression $sys =  \textsf{replace}(Sys, oP, nP)$ modifies $Sys$ by replacing inside it, the behaviour of $oP$ by the new behaviour expressed by $nP$.

Formally the channels shared by $Sys$ and $oP$ are renamed in $oP$ with a new name unused in $Sys$ and $nP$. Then $Sys$ is composed with $nP$.
Consequently if nc is the channel shared by the three processes, c a fresh channel, then we have:\\
\centerline{\textsf{replace}$(Sys, oldP, newP)$ = \textsf{compose}$(Sys, (Sys\selectP oP)$ \textsf{rename} nc \textsf{as} c, nP)}

\vspace{-0.2cm}
\paragraph{\textbf{\textsf{remove}: removing a process from a composition}}
 A given process can be removed from a system. The  \textsf{remove} operator (symbolically denoted by $\downarrow$) requires two arguments: a system $S$ made at least with two processes, a process $P$ already part of $S$. The process $P$  will be removed from $S$; this is symbolically denoted by $S\downarrow P$.
For instance the expression $sys =  \textsf{remove}(\textsf{compose}(P_1, P_2, P_3), P_2)$ results in a system composed of the processes $P_1$ and $P_3$.

\vspace{-0.2cm}
\paragraph{\textbf{\textsf{extractChan}: listing the channels of a process}}
This operator, when applied to a process, gives the list of channels used inside the process. The channels of processes can then be compared, reused, renamed, hidden.

\medskip
These operators constitute our target algebra and practically a core language: $\langle {\mathcal P}, \{\textsf{compose, replace, select, remove}\}, \{\textsf{rename, extractChan}\} \rangle$.
It is expressive enough, to describe the composition and the interaction between behavioural models as illustrated in the next section.

\vspace{-0.2cm}
\subsection{Illustration: a Heterogeneous Control System}

\begin{figure}
\centering
\begin{minipage}{.5\textwidth}
  \centering
\includegraphics[width=0.7\linewidth]{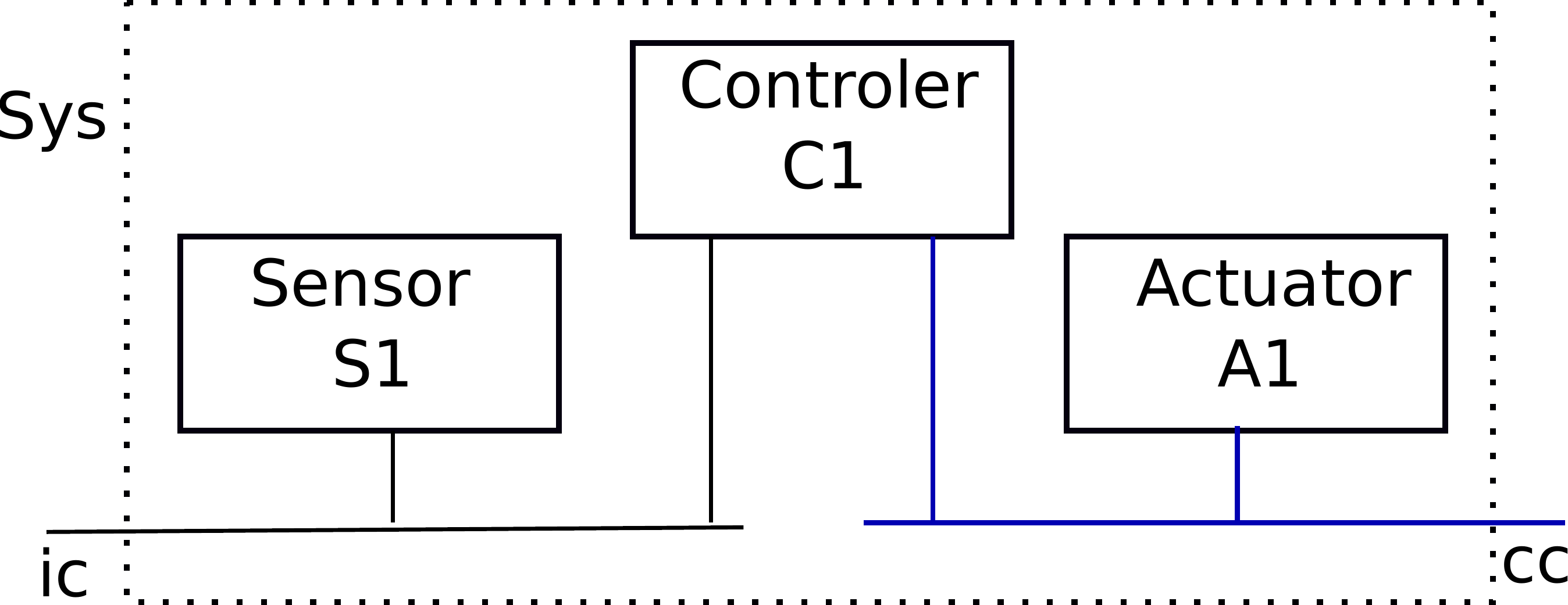}
 \captionof{figure}{Interaction net of processes}
\label{fig:processNet1}
\end{minipage}%
\begin{minipage}{.5\textwidth}
\centering
\includegraphics[width=0.9\linewidth]{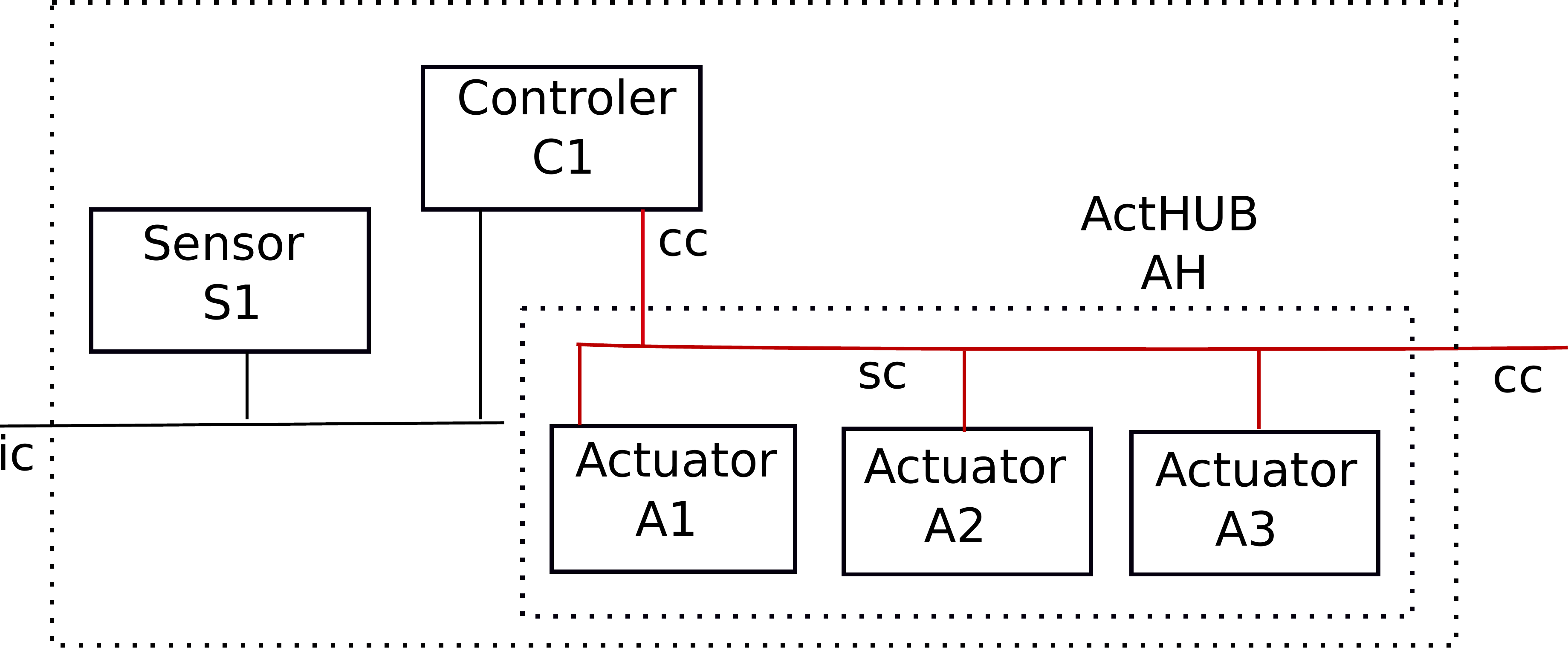}
 \captionof{figure}{Reshaped net of processes}
\label{fig:processNet2}
\end{minipage}
\end{figure}

We consider the interaction between a net of processes modelling a control system equipped with sensors, actuators and controlers. These devices come with their different behaviours,  from various vendors. From our heterogeneous modelling point of view the interfaces of the models are defined textually as follows and can be viewed as depicted in Fig.~\ref{fig:processNet1}. Note that the precise behaviours of the devices are not required at this stage. The communication channels at the interfaces are enough to  define the interaction.

\begin{center}
\begin{boxedminipage}{7.0cm}
\begin{tabular}{ll}
\textsf{Process Sensor[channel ic]()} \{...\} &S1,\\
\textsf{Process Controler[channel ic, cc]()} \{...\}~ & C1,\\
\textsf{Process Actuator[channel cc]()} \{...\} & A1.
\end{tabular}
\end{boxedminipage}
\end{center}

\medskip

Let a process controller $C_1$ interacting by reading a channel \textsf{ic} and writing on a control channel \textsf{cc}. Let a process Sensor $S_1$ interacting by sending data on the same channel \textsf{ic}.
At the modelling level, we could  simply write
$\textsf{compose}(C_1, S_1)$ so that $C_1$ and $S_1$ interact via \textsf{ic}. 
Considering $A_1$ as the actuator process interacting by reading the channel \textsf{cc}, then the description $sys = \textsf{compose}(\textsf{compose}(C_1, S_1), A_1)$ builds a new system (see Fig. \ref{fig:processNet1}) where the three processes interact together through the channels  \textsf{ic} and \textsf{cc}. The controler $C_1$ may send orders to the actuator $A_1$ depending on data read from $S_1$.
Now, we define two actuator processes $A_2$ and $A_3$ and  a hub of actuators $HA$ 
which sends its data to  $A_1$, $A_2$ and $A_3$:

\begin{center}
\begin{boxedminipage}{7.8cm}
\begin{tabular}{l}
Actuator A2\\
Actuator A3\\
HA = \textsf{compose}((A1 \textsf{rename}~ cc~ \textsf{as}~ sc), A2, A3)
\end{tabular}
\end{boxedminipage}
\end{center}


The behaviour expressed by $\textsf{replace}(sys,~ A_1,~ (HA~ \textsf{rename}~ sc~ \textsf{as}~ cc))$ results in a new system (see Fig. \ref{fig:processNet2}) where the controler $C_1$ is not anymore directly connected to $A_1$ but to $HA$ via the channel $cc$, a renaming of the previous $sc$ channel of $HA$. 
In the same way we can easily add new sensors $S_2, S_3$ into an existing pool of sensors with the compose operator: $\textsf{compose}(sys_1, S_2, S_3)$. The three sensors will write on the channel \textsf{ic}.



\vspace{-0.2cm}
\subsection{Semantics of Interaction}

The common interaction level considered here is related to the communications between the composed processes. As introduced in Section \ref{section:materials}, a compatibility domain is required to ensure consistency. Moreover, to reach this compatibility domain, from heterogeneous models, some filters could be applied. Typically compatible models have various additional specific facets.
For instance dealing with the domain of transition systems, some models may be structured according to multiple layers related to: communication channels, time properties, time constraints, QoS, $\cdots$.

Consequently, the models should be filtered according to these specific facets (time, channels, properties, etc) in order to get the desired compatibility and interaction level. At this stage we consider only the communication facet. Therefore the interaction between composed models are based on their communication through the used channels. The other facets will be consider latter in our work.

As far as the interaction between the behavioural models is considered, each process evolves according to the channels it uses. Interaction is based on communication via shared channels using emission and reception mechanisms (message passing).
Synchronous channels involve handshake communications. A reception takes place when a process applies the appropriate reception primitive relatively to a channel and, there is a (abstract) data sent on the addressed channel by the emission primitive applied by another process.
In the case of asynchronous channel, if there is nothing on the channel, the attempt of reception is aborted. 

\vspace{-0.2cm}
\subsection{Extending the Core Operators}

The previously defined core operators can be extended to compute other heterogeneous processes and systems. But more specifically, analysis operators should be defined to manipulate the built systems. 

Consider that we have a system made of sensors, actuators, controlers  and many other smart devices, making an adhoc network of communicating processes. We would like to plug a new device in the system so that it can interact with the existing processes; for instance a new plugged sensor detects the existing controlers and sends data to them, or a new plugged actuator joins the system and becomes ready to interact with the existing processes which send orders to the actuators.
But, it is not reasonably feasible to directly experiment with the system prior to the consistency decision.
Therefore reasoning at model level is appropriate.
The system builder may evaluate the forthcoming system, decide if some components or operations are correct or not before performing them on the real system. This is profitable if the used models and operations on models are trustworthy.
Consequently we would like to easily check the consistency of the new composition of processes prior to implementing it. For instance, with the aim of getting a diagnosis (at least True or False) of the envisaged composition, we would like to write the following:
\begin{center}
\begin{boxedminipage}{7.8cm}
\begin{tabular}{l}
\textsf{check}$(compose(sys, newSensor))$\\
\textsf{check}$(replace(sys, oldProcess, newProcess))$
\end{tabular}
\end{boxedminipage}
\end{center}

These scenario motivate the need to define the \textsf{check} operator which is not a process composition operator but an analysis one. Typically this kind of operators should implement at least the interaction compatibility, the absence of deadlock, liveness property. In the current stage of the work we reuse for this purpose, the existing tools of process algebra: \textsc{Uppaal} \cite{uppaal-tutorial04} which has its own graphical input description formalism, \textsf{SPIN} \cite{Holz97} which has the \textsf{Promela} language as input process description language and  \textsf{CADP} \cite{cadp2011} which uses  \textsf{Lotos} as input process description language.

\vspace{-0.2cm}
\section{Practical Analysis of Heterogeneous Models}
\label{section:experimentation}
%
We report on a case study dedicated to a distributed control system.  The system consists of a set of robots which supervise a geographically widespread area and take actions with respect to events in the area: intruders or found unusual objects.
Sub-components of the system are responsible of patrolling in different parts of the area and looking for preassigned objects; in case of detection of such objects a signal is sent to a supervisor.
Other sub-components follow a specific object or a detected intruder and communicate its location to the supervisor.

This kind of system is representative of systems being designed or used to assist people in hospitals, homes, remote care, restaurants, smart future factories, etc.
Indeed when constructing or maintaining large distributed systems, it is easy to collect components everywhere and combined them to build the systems. But when reliability should be tackled, we have to consider at least the models of the various components. These models are merely built with the appropriate description models instead of a common one for all the components. Hence an heterogeneity of the models at concrete or abstract level. 

As far as  our case study is concerned, at the design level, if we focus on the behavioural aspect related to the interaction between the system and its components, the system  can be modelled as the combination of several processes; each one having  specific features and a specific behaviour.

The target global system can be built incrementally with various components provided that the requirements are known. It is what we have experimented with. 
We consider the following components, which were built by considering the requirements of the case study.

A component, routine \textbf{patroller}, modelled by a process (\textsf{RPatrol}) which, after a connection to a supervisor, sends every $\delta$ seconds its collected data to the supervisor. This component was modelled using \textsf{Uppaal} because we need time feature provided by \textsf{Uppaal}.

A \textbf{scrutineer} component, modelled by a process  (\textsf{Scrutineer}) which, after a connection to a supervisor, gets from the supervisor its roadmap made of target objects to look at,  and then moves around and when one of the target object is found, it sends a signal to the supervisor. We modelled this component using \textsf{Promela} which provides high expressivity for handling data and control.

A \textbf{data collector} component (\textsf{dataCollector}) which, after a connection to a supervisor, becomes ready and sends on demand its collected data to the supervisor. This component is modelled as a simple transition system with the \textsf{DOT} formalism, seen as a common language used by several transition system formalisms. We give an excerpt in Fig.~\ref{fig:datacollector} for illustration purpose. Note the use of the synchronisation actions \textsf{connection} and \textsf{readState}, which will be renamed latter using the related operator of our kernel. As depicted in Fig.~\ref{fig:experimentNet2} (the process named \texttt{dtcltr1}) we can see the process as systematically renamed and translated into \textsf{Uppaal} by the tools we have built.

For experimentation purpose, we  built several modules to implement the core operators from the \textsf{DOT} format: channel renaming, process selection, process removing and  specific facet filtering (time, ...).

A \textbf{follower} component is dedicated to detect and follow intruder or assigned object. This component first try a connection to the supervisor; on success it scruts around and then follows the assigned object. The relative positions are then sent to the supervisor until the followed object leaves or when the supervisor orders to stop. We have specified this component as a \textsf{LOTOS} process.

\begin{figure}[!ht]
\begin{center}
{\small
\begin{boxedminipage}{6.5cm}
\begin{alltt}
digraph p_dataCollectorBot \{
 initSCRT -> S2 [label={\red connection!}];
 S2 -> initSCRT [label=KO?];
 S2 -> S3 [label=OK?];
 S3 -> S4 [label=ready!];
 S4 -> initSCRT [label=stop?];
 S4 -> S5 [label={\blue readState!}];
 S5 -> S4 [label=getState?];
\}
\end{alltt}
\end{boxedminipage}
}
\caption{A LTS model of the data collector in the DOT format}
\label{fig:datacollector}
\end{center}
\end{figure}

A supervisor or \textbf{global controler},  modelled as a process is written in \textsf{Uppaal}.

\subsection{Combining the Heterogeneous Processes}
\begin{figure}[!ht]
\begin{center}
\includegraphics[width=0.99\linewidth]{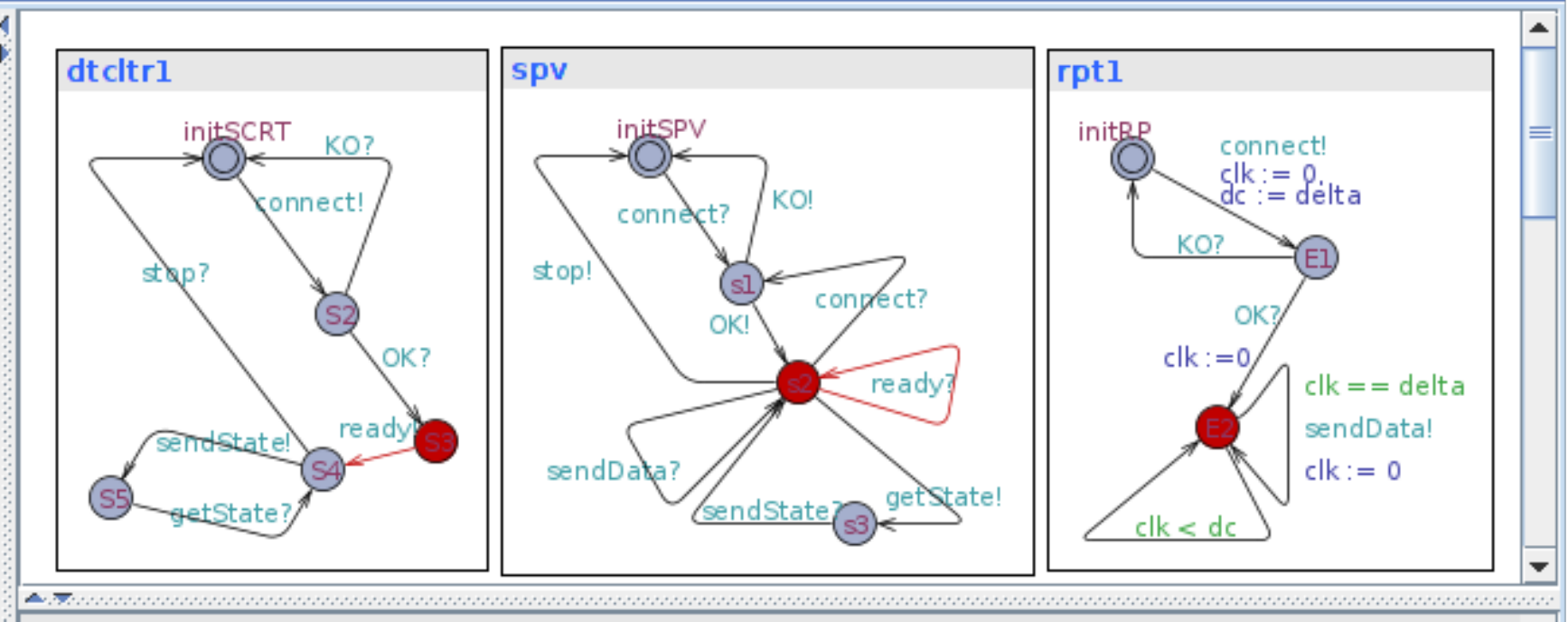}
 \caption{Interaction between composed processes of the system}
\label{fig:experimentNet2}
\end{center}
\end{figure}

We use the proposed operators to experiment with the composition of the previous processes. For instance Fig.~\ref{fig:experimentNet2} depicts the composition resulting from the term:
\textsf{compose}($dtctrl1$ \textsf{rename} connection as connect, $spv$, $rpt1$)
where:
\begin{itemize}
\item $dtctrl1$ is an instance of the process \textsf{dataCollector} after two renamings and a translation into  \textsf{Uppaal}: \\
\centerline{((\textsf{dataCollector} \textsf{rename} connection as connect) \textsf{rename} readState as sendState)}
\item $rpt1$ is an instance of the patroller \textsf{RPatrol},
\item $spv$ is an instance of \textsf{supervisor}.
\end{itemize}

Several $\zeta_i$ morphisms were used to embed the provided models into LTS. The \textsf{DOT} formalism have been intensively used.
\begin{itemize}
\item $\zeta_{dot2upa}$  A translator from \textsf{DOT} to \textsf{Uppaal} using \textsf{XML} as the internal representation of the \textsf{Uppaal} tool. 
\item  $\zeta_{pml2upa}$  A translator from \textsf{promela} to \textsf{Uppaal}. We reuse the features of SPIN; from \textsf{Promela} we  generate a \textsf{DOT} format of a process using the features provided by \textsf{SPIN}. Then we reuse our module that embed  \textsf{DOT} into \textsf{Uppaal}.
\item $\zeta_{dot2lotos}$  A transator from \textsf{DOT} to \textsf{Lotos} process.
\end{itemize}

\medskip

The proposed algebra and method for composing heterogeneous processes enable us to deal with several examples.
In case of examples where the communications actions do not bear data the basic structure morphisms are easily implemented and work well.
In case where we have a complex structuring of data used as parameters of communication actions, the structure morphisms are not straightforward, since specific interpretations can be given to the data; an action such as \textsf{chan!value} may denote not simply the sending of  \texttt{value} on the channel but the sent value may also conveying a semantic such as a process identifier.

The various experiments also reveal that a more expressive formalism is preferable to be used as the target of embedding in order to avoid loosing information. For this purpose we target the \textsf{Lotos} language for further experiments.

\subsection{Analysis of the Global System}
Remind, one interest of gathering the various models is being able to perform an analysis of the global system.
In the current case of behavioural models,  as stated before, at least standard liveness properties should be analysed on the basis of the global system. We have been able to perform such analysis on the basis adopted for the composition.\\

\textit{Deadlock}. The absence or presence of a deadlock is a standard property. It is expressed in LTL logic provided by the \textsf{Uppaal} tool as follows:
\begin{center}
\verb|A[] not deadlock|
\end{center}

\textit{Progress}. Specific progress or reachability properties can be stated using the \textsf{Uppaal} model. For instance we check that both the data collector and the routine patroller interact simultaneously with the controler and can reach some given states of their behaviour.  This is expressed as follows:
\begin{center}
\verb| E<> dtctrl1.ready  and  rpt1.E2 |
\end{center}

\begin{figure}[!ht]
\begin{center}
\includegraphics[width=0.99\linewidth]{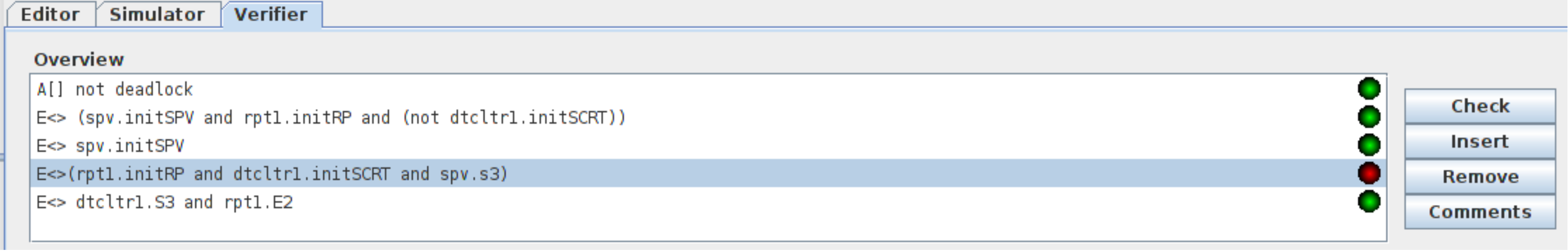}
 \caption{Examples of checked properties}
\label{fig:experimentChecking1}
\end{center}
\end{figure}

Examples of other properties we have checked are depicted in Fig.~\ref{fig:experimentChecking1}\\

\subsection{Tool Support: \textsf{aZiZa}}
To experiment with examples, as shown in the previous section, we have developed the main modules of a prototype tool called \textsf{aZiZa}\footnote{\url{aziza.ls2n.fr}}, to support our method proposal.
This is necessary to validate the proposed concepts and to improve the global composition method.  The core modules are described bellow.

\noindent
\begin{center}
{\small
\begin{tabular}{lp{8.5cm}}
\hline
Dot to Uppaal &  A translator  between an LTS formatted with DOT and the XML repremsentation of \textsf{Uppaal} processes\\
Promela to Uppaal & A translator from Promela to Uppaal. Here we  reuse the dot generator of the Spin tool. Then we combine with the previous translator.\\
Label Filtering &  This module is a filter to extract a part of a dot LTS\\
Channel renaming & This module is used to rename a channel of a process\\
Promela process splitting &  It splits a complete promela system into several processes. This module is developed for the purpose to reuse a given process. \\
Process selection &  This module achieves the selction of a target process from a given composition.\\
\hline
\end{tabular}
}
\end{center}

\medskip \medskip
Yet, all the steps are not automated. Some steps are still manual, for instance the result of the translation of the process, is to be copy-paste in a \textsf{Uppaal} file, or the graphical layout are not automated.

\begin{figure}[!ht]
\includegraphics[width=0.99\linewidth]{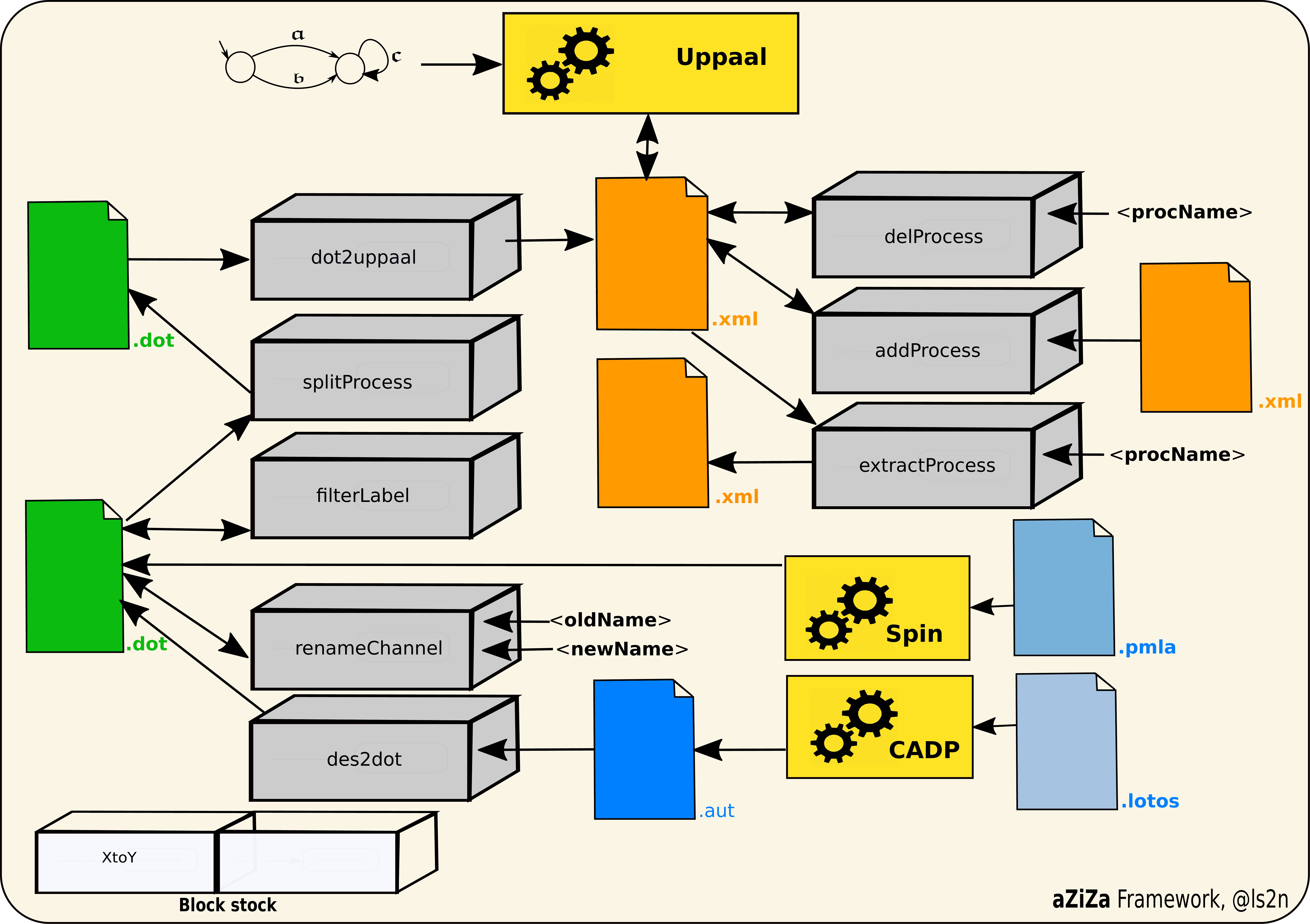}
 \caption{The architecture of the \textsf{aZiZa} tool}
\label{fig:toolArchitecture}
\end{figure}

The architecture of the prototype tool is depicted in Fig. \ref{fig:toolArchitecture}. Its is made of interconnected modules (the blocks in the figure).  As for now we intensively use intermediate files to connect the blocks in oder to make easy the independent development of the blocks.

\vspace{-0.3cm}
\section{Conclusion}
 \label{section:conclu}

We have shown that under the hypothesis of semantic domain compatibility, the composition of heterogeneous behavioural models can be overcomed. We have used labelled transition systems as common semantics domain to found our method of composition.
The method is based on an algebra of operators that focuses on the manipulation of channels which are the communication mechanism between the composed models.
We have equipped the method with tools in order to experiment with case studies.
We illustrated the article with a case study of a distributed control system where various processes cooperate to control objects evolving within a given area.
The case study reported here is a part of a series of experiments that serve as a mean of assessment of the proposal and a testbed to improve the tools under development. 

One representative of the related works is Ptolemy II \cite{EdLee2010,ptolemy-TripakisSSL13}.
Ptolemy achieves the interaction between different actor-oriented models using an abstract semantics (namely the actor semantics). 
It also enables the use of finite state machines in place of actor-oriented models, but the interaction works rather as a global scheduler, controling a sequential execution flow  of the FSM considered each as a global state linked via a port to another one. Moreover Ptolemy II is a general purpose heavy weight composition framework. Unlikely we  target a specific, flexible and extensible framework dedicated to the composition and analysis of behavioural models dedicated to the growing small aperture nets of processes.

There are some aspects that are not yet considered at this stage of the work; for example we have considered some components dealing locally with time constraints but dealing with time constraints at the global level is a challenge.
As future work, we have planned  experimentations with the more expressive CADP framework and especially its \textsf{exp.open} composition tool.
We plan some improvements among which the propagation of global properties inside local components and vice versa. For this purpose we are investigating the Property Specification Language \cite{psl2007}, an IEEE standard, as a pivotal for property passing through the components and through the various tools. 



\vspace{-0.2cm}

\bibliographystyle{abbrv}

\begin{thebibliography}{10}

\bibitem{psl2007}
{IEC 62531 Ed. 1 (2007-11) (IEEE Std 1850-2005): Standard for Property
  Specification Language (PSL)}.
\newblock {\em IEC 62531:2007 (E)}, pages 1--152, Dec 2007.

\bibitem{TS_Arnold93}
Andr{\'{e}} Arnold.
\newblock {Verification and Comparison of Transition Systems}.
\newblock In M{-}C. Gaudel and J{-}P. Jouannaud, editors, {\em TAPSOFT'93:
  Theory and Practice of Software Development, CAAP/FASE}, volume 668 of {\em
  LNCS}, pages 121--135. Springer, 1993.

\bibitem{uppaal-tutorial04}
Gerd Behrmann, Alexandre David, and Kim~G. Larsen.
\newblock {A Tutorial on {\sc Uppaal}}.
\newblock In M.~Bernardo and F.~Corradini, editors, {\em 4th International
  School on Formal Methods for the Design of Computer, Communication, and
  Software Systems, SFM-RT 2004}, number 3185 in LNCS, pages 200--236.
  Springer--Verlag, 2004.

\bibitem{systemC2010}
D.C. Black, J.~Donovan, and and A.~Keist B.~Bunton, editors.
\newblock {\em {SystemC: From the Ground Up, Second Edition}}.
\newblock Springer, 2010.

\bibitem{alfarohezinger2001interface}
Luca De~Alfaro and Thomas Henzinger.
\newblock {Interface Theories for Component-based Design}.
\newblock In {\em Embedded Software}, pages 148--165. Springer, 2001.

\bibitem{DBLP:journals/pieee/EkerJLLLLNSX03}
Johan Eker, J{\"o}rn~W. Janneck, Edward~A. Lee, Jie Liu, Xiaojun Liu,
  J.~Ludvig, Stephen Neuendorffer, S.~Sachs, and Yuhong Xiong.
\newblock {Taming Heterogeneity - the Ptolemy Approach}.
\newblock {\em Proceedings of the IEEE}, 91(1):127--144, 2003.

\bibitem{SysML-Friedenthal2015}
Sanford Friedenthal, Alan Moore, and Rick Steiner.
\newblock {A Practical Guide to SysML}.
\newblock The MK/OMG Press. Morgan Kaufmann, Boston, 2015.

\bibitem{cadp2011}
Hubert Garavel, Fr{\'{e}}d{\'{e}}ric Lang, Radu Mateescu, and Wendelin Serwe.
\newblock {{CADP} 2011: a Toolbox for the Construction and Analysis of
  Distributed Processes}.
\newblock {\em {STTT}}, 15(2):89--107, 2013.

\bibitem{GoderisBALG09}
Antoon Goderis, Christopher~X. Brooks, Ilkay Altintas, Edward~A. Lee, and
  Carole~A. Goble.
\newblock {Heterogeneous Composition of Models of Computation}.
\newblock {\em Future Generation Comp. Syst.}, 25(5):552--560, 2009.

\bibitem{GordonHOL}
M.J.C. Gordon.
\newblock {\em {Introduction to HOL: A Theorem Proving Environment}}.
\newblock Cambridge University Press, 1993.

\bibitem{Holz97}
G.~J. Holzmann.
\newblock {The Spin Model Checker}.
\newblock {\em IEEE Transactions on Software Engineering}, 23(5):279--295, May
  1997.

\bibitem{EdLee2010}
Edward~A. Lee.
\newblock {Disciplined Heterogeneous Modeling}.
\newblock In D.~C. Petriu, N.~Rouquette, and {\O}.~Haugen, editors, {\em 13th
  International Conference, {MODELS} 2010, Oslo, 2010}, volume 6395 of {\em
  LNCS}, pages 273--287. Springer, 2010.

\bibitem{ISOlotos88}
LOTOS.
\newblock {\em {A Formal Description Technique Based on The Temporal Ordering
  of Observational Behaviour}}.
\newblock IOS - OSI, Geneva, 1988.
\newblock International Standard 8807.

\bibitem{Milner92}
R.~Milner, J.~Parrow, and D.~Walker.
\newblock {A Calculus of Mobile Processes}.
\newblock {\em Journal of Information and Computation}, 100, 1992.

\bibitem{Mil89}
Robin Milner.
\newblock {\em {Communication and Concurrency}}.
\newblock Prentice-Hall, 1989.

\bibitem{DBLP:RoscoeD11}
A.~W. Roscoe and J.~Davies.
\newblock {{CSP} (Communicating Sequential Processes)}.
\newblock In David~A. Padua, editor, {\em Encycl. of Parallel Computing}.
  Springer, 2011.

\bibitem{Roscoe98}
A.W. Roscoe.
\newblock {\em {The Theory and Practice of concurrency}}.
\newblock Prentice-Hall, 1998.

\bibitem{RothKinney_Mealy_2004}
C.~H. Roth and L.~L. Kinney.
\newblock {\em {Fundamentals of Logic Design}}.
\newblock Thomson, 2004.

\bibitem{ptolemy-TripakisSSL13}
S.~Tripakis, C.~Stergiou, C.~Shaver, and E.~A. Lee.
\newblock {A Modular Formal Semantics for Ptolemy}.
\newblock {\em Mathematical Structures in Computer Science}, 23(4):834--881,
  2013.

\end{thebibliography}

\end{document}